\def\FIGUREPATH{./}
\def\RESOURCEPATH{./}

\documentclass[conference,9pt]{IEEEtran}
\usepackage{graphicx, amsfonts}
\usepackage{amssymb,color,booktabs}
\usepackage{amsmath}
\usepackage{tabularx}
\usepackage{url}

\def\BibTeX{{\rm B\kern-.05em{\sc i\kern-.025em b}\kern-.08em
		T\kern-.1667em\lower.7ex\hbox{E}\kern-.125emX}}

\title{Distributed Microphone Speech Enhancement based on Deep Learning\\
}

\author{\IEEEauthorblockN{1\textsuperscript{st} Syu-Siang Wang}
	\IEEEauthorblockA{\textit{Research Center for Information Technology Innovation} \\
		\textit{Academia Sinica}\\
		Taipei, Taiwan\\
		sypdbhee@citi.sinica.edu.tw}
	\and
	\IEEEauthorblockN{2\textsuperscript{nd} Yu-You Liang}
	\IEEEauthorblockA{\textit{Department of Electrical Engineering} \\
		\textit{Yuan Ze University}\\
		Taoyuan, Taiwan \\
		y.y.liang.hchs@gmail.com}
	\and
	\IEEEauthorblockN{3\textsuperscript{rd} Jeih-weih Hung}
	\IEEEauthorblockA{\textit{Department of Electrical Engineering} \\
		\textit{National Chi Nan University}\\
		Nantou, Taiwan \\
		jwhung@mail.ncnu.edu.tw}
	\and
	\IEEEauthorblockN{4\textsuperscript{th} Yu Tsao}
	\IEEEauthorblockA{\textit{Research Center for Information Technology Innovation} \\
		\textit{Academia Sinica}\\
		Taipei, Taiwan\\
		yu.tsao@citi.sinica.edu.tw}
	\and
	\IEEEauthorblockN{5\textsuperscript{th} Hsin-Min Wang}
	\IEEEauthorblockA{\textit{Institute of Information Science} \\
		\textit{Academia Sinica}\\
		Taipei, Taiwan\\
		whm@iis.sinica.edu.tw}
	\and
	\IEEEauthorblockN{6\textsuperscript{th} Shih-Hau Fang}
	\IEEEauthorblockA{\textit{Department of Electrical Engineering} \\
		\textit{Yuan Ze University}\\
		Taoyuan, Taiwan\\
		shfang@saturn.yzu.edu.tw}
}

\begin{document}
%
\maketitle
\begin{abstract}
Speech-related applications deliver inferior performance in complex noise environments. Therefore, this study primarily addresses this problem by introducing speech-enhancement (SE) systems based on deep neural networks (DNNs) applied to a distributed microphone architecture, and then investigates the effectiveness of three different DNN-model structures. The first system constructs a DNN model for each microphone to enhance the recorded noisy speech signal, and the second system combines all the noisy recordings into a large feature structure that is then enhanced through a DNN model. As for the third system, a channel-dependent DNN is first used to enhance the corresponding noisy input, and all the channel-wise enhanced outputs are fed into a DNN fusion model to construct a nearly clean signal. All the three DNN SE systems are operated in the acoustic frequency domain of speech signals in a diffuse-noise field environment. Evaluation experiments were conducted on the Taiwan Mandarin Hearing in Noise Test (TMHINT) database, and the results indicate that all the three DNN-based SE systems provide the original noise-corrupted signals with improved speech quality and intelligibility, whereas the third system delivers the highest signal-to-noise ratio (SNR) improvement and optimal speech intelligibility.
\end{abstract}
\section{Introduction}\label{sec:intro}
Real-world environments are always contain stationary and/or time-varying noises that are received together with speech signals by recording devices \cite{michelsanti2017conditional}. The received noises inevitably degrade the performance of multi-channel (MC)-based human--human and human--machine interfaces, and this issue has attracted significant attention over the years \cite{8683662,bitzer2001multi,8683732,qi2020tensor,bitzer1999multimicrophone,liu1995room,grais2018raw}. In recent decades, numerous MC speech-enhancement (SE) approaches have been proposed to alleviate the effect of noise and improve the quality and intelligibility \cite{de2017adaptive,bagheri2019exploiting,8683768,1255457,heymann2016neural,chen2018building} of received speech signals. In general, most of these approaches have been proposed for use in a microphone array architecture, wherein multiple microphones are compactly placed in a small space. For example, the beam-forming algorithm, one of the most popular methods that exploit the spatial diversity of received signals to design a linear filter in the frequency domain, aims to preserve the signal received from the target direction while attenuating noise and interference from other directions \cite{hoshuyama1999robust,emura2018distortionless,chang2019mimo}. Recently, several novel approaches have combined deep-learning-based algorithms, such as deep neural networks (DNNs) and convolutional neural networks (CNNs) \cite{wang2018supervised}, with the beam-forming process to further promote the enhanced capability of an MC SE system \cite{zhang2017deep,wang2018all,tan2019real,araki2015exploring} so as to produce speech signals with even higher quality. In addition to beam-forming-based approaches, in \cite{tawara2019multi}, the multiple recordings are directly enhanced in the time domain along the specified spatial direction through a denoising auto-encoder, and this method benefits automatic speech recognition (ASR) systems by reducing recognition errors. 

In contrast, some researchers pay more attention on performing SE on the architecture of distributed-microphones \cite{liu2019multichannel,hassani2017real,Matheja2013,li2018distributed,TU201596}. This physical configuration, consisting of many individual self-powered microphones or microphone arrays, can be deployed in a large area \cite{7503169,6101302}. Therefore, one or more received signals with a higher signal-to-noise ratio (SNR) and a direct-to-reverberant ratio can be used for the distributed-microphone enhancement system to achieve better sound quality and intelligibility. 

In general, a fusion center (FC) and a distributed signal processor (also called ad-hoc) are two alternative forms used in the distributed-microphone architecture \cite{markovich2015optimal}. For the FC, each recording device can transmit the recorded sounds to a powerful central processor that aims at reconstructing nearly clean speech relative to the selected target speaker. Some successful approaches associated with this architecture include robust principal component analysis \cite{li2018distributed}, generalized eigen value decomposition \cite{markovich2009multichannel}, MC Wiener filter \cite{serizel2014low}, and optimal MC frequency domain estimators \cite{TRAWICKI2012345}. In comparison, the ad-hoc-based enhancement system, however, enhances the noisy input locally in the individual device and then shares the result with its neighbors for further refinement. Some well-known techniques of this type include distributed linearly constrained minimum variance \cite{bertrand2013distributed}, linearly constrained distributed adaptive node-specific signal estimation \cite{bertrand2011distributed}, distributed generalized sidelobe canceler \cite{markovich2012distributed}, and distributed maximum signal to interference-plus-noise filtering \cite{tavakoli2017distributed}.

In this study, three novel SE systems based on DNNs are introduced and investigated for the distributed-microphone architecture. For the first system, we train the DNN model for each microphone channel to enhance the corresponding noisy recordings. In other words, a large MC SE system is divided into several single-channel noise reduction tasks, and this system is called ``DNN--S''. Next, the second system follows a process similar to the work in \cite{tawara2019multi}, wherein the multiple noisy utterances received are transmitted to an FC and then aggregated and used as input to a DNN model for producing the final enhanced signals. Because an FC is used here, this system is called ``DNN--F''. Finally, the third system comprises two operational stages. The first stage acts locally in each channel device by enhancing the noisy input with a DNN, and in the second stage, all the enhanced local-channel signals are combined and further processed with a fusion DNN. We call this system ``DNN--C'' because it nearly combines the two previous systems to facilitate the enhancement. Notably, in our opinion the main contribution of this study is two-fold: First, we show that the deep learning-based algorithm can be successfully applied to the distributed microphone architecture and reveals its promising capability in reducing noise effect. Second, three types of deep neural network (DNN) structures are compared with each other when they are adopted in the distributed-microphone SE task, and we investigate the underlying reasons for the corresponding performance difference.

The rest of this paper is organized as follows. Section \ref{sec:threemodel} introduces the aforementioned three systems, namely DNN--S, DNN--F, and DNN--C. Experiments and the respective analysis are given in Section \ref{sec:experi}. Section \ref{sec:conclu} presented conclusions and a future avenue.

\section{Three different multi-channel enhancement system}\label{sec:threemodel}
\begin{figure}[!t]
{\centering \includegraphics[width=\columnwidth]{\FIGUREPATH 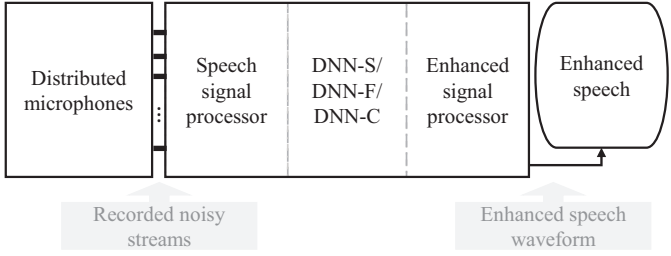}
\caption{Block diagram of a DNN-based speech-enhancement system with a distributed-microphone architecture.}\label{fig:ovall}}
\end{figure}
Figure \ref{fig:ovall} shows the general diagram of a distributed-channel SE architecture common to the three presented systems. The original clean signal shown in this figure is first corrupted by the background diffuse noise and/or reverberation and is received by the distributed microphones. Next, the short-time Fourier transform (STFT) and logarithmic operation are performed on the received signals to obtain the log-power magnitude spectra (LPS) in the speech signal processor block. The MC SE system based on DNN, DNN--S, DNN--F, or DNN--C, is then used to generate the enhanced version from the noisy LPS input. Finally, in the enhanced signal processor block, the inverse STFT (ISTFT) is applied to the enhanced LPS, together with the original phase component preserved from the specific channel to provide the final enhanced speech waveform. In the following three sub-sections, we will introduce the detail process of the aforementioned three DNN-based SE systems. 

\subsection{DNN--S enhancement systems}
\begin{figure}[!t]
{\centering \includegraphics[scale=0.95]{\FIGUREPATH 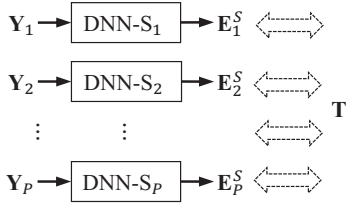}
\caption{Block diagram of the DNN--S SE system.}\label{fig:dnns}}
\end{figure}

For the DNN--S architecture as shown in Fig. \ref{fig:dnns}, a DNN, presented by DNN--S$_p$, was trained for each of the $P$--microphone channels using the channel-wise noisy speech feature set $\mathbf{Y}_p$ with the respective clean counterpart $\mathbf{T}$, where $P$ is the total number of microphone channels, and $p$ is the channel index. Consider an $L$--layer DNN--S$_p$, an arbitrary $l$th layer is formulated in Eq. \eqref{eq:dnns} in terms of the input-output relationship ($\mathbf{z}^{(l-1)}$, $\mathbf{z}^{(l)}$):
\begin{equation}\label{eq:dnns}
\mathbf{z}^{(l)}=\sigma^{(l)}\left(h^{(l)}(\mathbf{z}^{(l-1)})\right),\quad l=1,\cdots,L,
\end{equation}
where $\sigma^{(l)}$ and $h^{(l)}$ are the ReLU activation and linear transformation functions, respectively. Notably, the input and output layers correspond to the first and $L$-th layers, respectively. In addition, for DNN--S$_p$, we have $\mathbf{z}^{(0)}=\mathbf{Y}_p$ and $\mathbf{z}^{(L)}=\mathbf{E}_p^S$, where $\mathbf{E}_p^S$ is the ultimate output of this system. The DNN parameters are obtained by means of supervised training that minimizes the mean squared error (MSE) between $\mathbf{E}_p^S$ and the noise-free counterpart $\mathbf{T}$. 

\subsection{DNN--F enhancement system}
\begin{figure}[!t]
{\centering \includegraphics[scale=0.95]{\FIGUREPATH 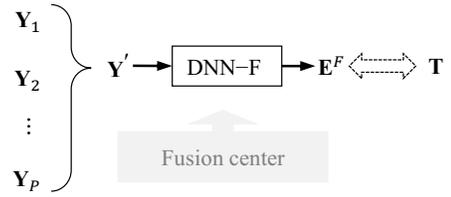}
\caption{Block diagram of the DNN--F speech-enhancement system.}\label{fig:dnnf}}
\end{figure}
Figure \ref{fig:dnnf} illustrates the DNN--F block diagram. According to this figure, we collect the channel-wise noisy features, $\mathbf{Y}_p$, $p=1, 2, ..., P$ from all of the $P$--microphone channels, and concatenate them to form a long feature $\mathbf{Y}'_p$ in a frame-wise manner, i.e., $\mathbf{Y}'=[\mathbf{Y}_1;\cdots,\mathbf{Y}_p;\cdots,\mathbf{Y}_P]$. Then, the long features in the training set are used to train a DNN model in FC, which can be formulated as follows 
\begin{equation}\label{eq:dnnf}
\mathbf{E}^F=\mbox{\textit{DNN-F}}\{\mathbf{Y}'\},
\end{equation}
where $\mbox{\textit{DNN--F}}\{\cdot\}$ represents the operation of the used DNN model, and $\mathbf{E}^F$ is the corresponding enhanced output. DNN--F applies a fusion model to enhance the noisy features from all channels concurrently, whereas DNN--S exploits $P$-channel-dependent models.

\subsection{DNN--C enhancement system}
\begin{figure}[!t]
{\centering \includegraphics[width=0.9\columnwidth]{\FIGUREPATH 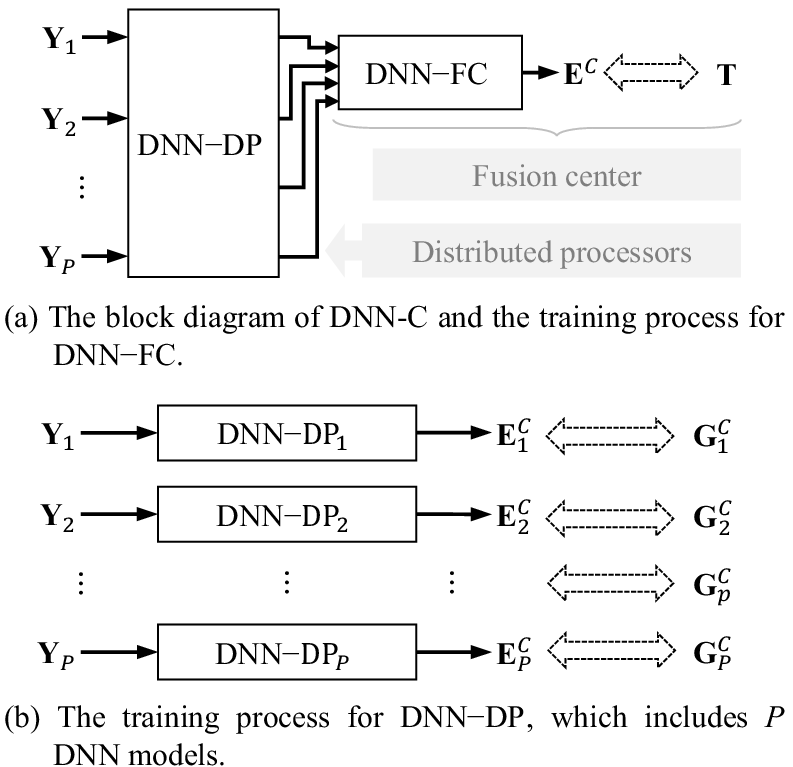}
\caption{DNN--C block diagram depicted in (a) contains both DNN--FC and DNN--DP functions. The training target for DNN--FC is the clean LPS $\mathbf{T}$, while that for DNN--DP is the ground truth voice $\mathbf{G}_p^C$. In addition, the DNN--DP model is determined first in (b), and then fixed for performing DNN--FC in (a).}\label{fig:dnnc}}
\end{figure}

The third proposed system, DNN--C, consists of two stages, the distributed processing stage and the fusion stage. Both stages use DNN models and are therefore represented by ``DNN--DP'' and ``DNN--FC'', respectively. The general diagram of DNN--C is shown in Fig. \ref{fig:dnnc} (a), and the detailed configuration of the DNN--DP stage is shown in Fig. \ref{fig:dnnc} (b). According to Fig. \ref{fig:dnnc} (b), $P$-self-powered devices, each for every individual microphone channel, are employed and a channel-specific DNN model denoted by DNN--DP$_p$ is conducted on each of these devices for suppressing noise from the input $\mathbf{Y}_p$ to produce the enhanced features, denoted by $\mathbf{E}_p^C$. Like DNN--S and DNN--F models, each DNN--DP$_p$ is composed of $L$ layers with ReLU activation and linear transformation functions, and can be formulated as follows:
\begin{equation}\label{eq:dnndp}
\mathbf{E}_p^C=\mbox{\textit{DNN-DP}}_p\{\mathbf{Y}_p\}.
\end{equation}
Therefore, the $P$--DNN models---DNN--DP$_1$, DNN--DP$_2$,$\cdots$, DNN--DP$_P$---are first estimated in the DNN--DP stage. In particular, the target feature for each of the $P$--DNN models, denoted by $\mathbf{G}_p^C$, is created by the channel-dependent clean speech, which is recorded from the $p$th microphone in the noise-free environment. That is, the $\mathbf{Y}_p$--$\mathbf{G}_p^C$ speech pair is used to train the associated DNN--DP$_p$ model.

As for the second stage, ``DNN--FC'', the $P$--DNN outputs at the first stage are concatenated to form a new feature $\mathbf{E}^{C'}$, i.e., $\mathbf{E}^{C'}=[\mathbf{E}_1^{C};\cdots;\mathbf{E}_p^{C};\cdots;\mathbf{E}_P^{C}]$, that is used together with the clean target $\mathbf{T}$ to train the DNN--FC model, which can be formulated by:
\begin{equation}\label{eq:dnnfc}
\mathbf{E}^C=\mbox{\textit{DNN-FC}}\{\mathbf{E}^{C'}\}.
\end{equation}
Therefore, the overall operation of the DNN--C system can be represented by
\begin{equation}\label{eq:dnnc}
\mathbf{E}^C=\mbox{\textit{DNN-FC}}\{\textit{DNN-DP}_1\{\mathbf{Y}_1\},\cdots, \textit{DNN-DP}_P\{\mathbf{Y}_P\}\}.
\end{equation}

\subsection{Phase component}
Briefly, the MC SE systems presented here enhance multiple signal sources in the frequency domain. Therefore, an ISTFT is applied to the updated spectrogram to produce the enhanced time-domain signal. It is worth mentioning that the phase component in the enhanced spectrogram is generated from a noisy speech through STFT from a specific channel. For DNN--S, the enhanced signal processor shown in Fig. \ref{fig:ovall} is performed individually on each channel for reconstructing the speech waveform with the enhanced LPS and the preserved noisy phase. Conversely, the phase used for DNN--F and DNN--C is extracted from one of the noisy channels, where it is assumed that the recorded sound has the highest SNR and optimal speech quality among all channels. 

\section{Experiments}\label{sec:experi}
In the following subsections, we first describe the experimental setup of the SE distributed-microphone tasks and then present the experimental results together with some discussions for the presented systems.

\subsection{Experimental setup}
\begin{figure}[!t]
{\centering \includegraphics[scale=0.9]{\FIGUREPATH 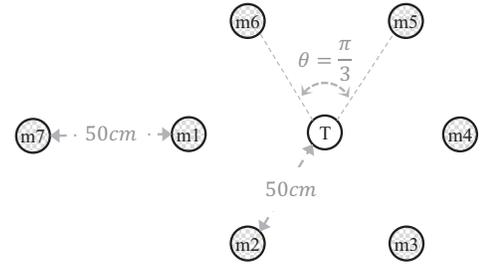}
\caption{MC system consists of seven microphones (``m1'', ``m2'',$\cdots$, ``m7''); microphones m1, m2,$\cdots$, m6 were placed around the speaker T with a radius of 0.5 meter, and m7 was put behind m1 and oriented towards T at a distance of 1 meter.}\label{fig:conf}}
\end{figure}
The layout of the distributed MC system is shown in Fig. \ref{fig:conf}. Seven microphones ($P=7$) of the same brand and model (Sanlux HMT-11) were used and denoted by ``m1'', ``m2'',$\cdots$, ``m7'', respectively, while the target speaker was represented by ``$T$''. In this system, six microphones, m1, m2,$\cdots$, m6, were placed around the speaker and equally spaced by an angle $\pi/3$ in the median plane with a 0.5-meter radius. The m7 microphone was placed right behind the m1 microphone and oriented towards the $T$ speaker at 1 meter. Notably, recording speech signals in such a distributed MC configuration might introduce different amounts of time delays across microphone channels \cite{redondi2009geometric}. However, here we assume that the difference in time delay across channels is small and can be negligible for simplicity in analysis.

For the evaluation task, we used the Taiwan Mandarin Hearing in Noise Test (TMHINT) \cite{huang2005development} to prepare the speech dataset. According to the script provided by the TMHINT dataset, 300 phrases were selected as the training set, while the remaining 20 utterances were used for testing. Phrases in the training set were individually pronounced by a male and a female in a noise-free environment at a sampling rate of 16 kHz. These recordings were then corrupted by eight different types of noise (cockpit, machine gun, alarm, cough, PC fan, pink, babble, and engine) at eight different noise levels (ranging from $-5$ to $16$ dB SNRs with a 3 dB interval). Thus, 38,400 utterances ($300\times 2\times 8\times 8$) were reproduced and then received by each of the seven microphones. On the other hand, the test utterances were first recorded by another male and female speaker and then contaminated with siren and street noises at -5, 0 and 5 dB SNRs. Therefore, there were 240 noisy utterances ($20\times 2\times 2\times 3$) transmitted to each of the seven microphones. 

For any of the seven microphone channels, each received utterance was first split into overlapped frames with a 32-ms frame duration and a 16-ms frame shift, and a series of 257-dim frame-wise LPS were constructed accordingly. The context feature for each frame was then created by concatenating the LPS of three neighboring frames. Therefore, the dimensions of the input feature were 771 ($257\times 3$) for each DNN--S$_p$, 771 ($257\times 3$) for DNN--DP$_p$, 5397 ($257\times 3\times 7$) for DNN--F, and 1799 ($257\times 7$) for DNN--FC, respectively. By contrast, the output dimensions of DNN--S$_p$, DNN--F, DNN--DP$_p$, and DNN--FC were 257. In addition, each DNN--S$_p$ and DNN--F model consisted of seven layers and 2,048 nodes per hidden layer. A DNN--DP$_p$ model was arranged to have five layers, whereas the DNN--FC model had four layers. The number of nodes for each hidden layer of DNN--DP and DNN--FC models was set to 2,048.

Three metrics were used to evaluate the enhanced utterances, including perceptual evaluation of speech quality (PESQ) \cite{rix2001perceptual}, short-time objective intelligibility (STOI) \cite{taal2011algorithm}, and segmental SNR improvement (SSNRI) \cite{mermelstein1979evaluation,plapous2006improved}. Higher scores for PESQ, STOI, and SSNRI indicate better enhanced performance.

\subsection{Experimental results} 
The averaged STOI scores of the test utterances (input) and enhanced utterances (output) of individual microphone channels of the DNN--S system are listed in Tables \ref{tab:stoinoidnns}. It is clear that almost all DNN--S$_p$ models could improve the STOI score, except for channel m1 in DNN--S$_1$. One possible reason is the overfitting issue that may have degraded the generalization capability of the model in the testing environments. Meanwhile, the STOI scores of noisy testing utterances varied with different recording channels, which may be owing to the different locations of the microphones deployed in a space, as shown in Fig. \ref{fig:conf}, despite the fact that these microphones were at the same distance from the speaker. In addition, Table \ref{tab:sdinoidnns} lists the SSNRI scores achieved by individual DNN--S$_p$ models. Similar to the case in Table \ref{tab:stoinoidnns}, all the channel-wise DNN--S models bring significant SNR improvements, whereas DNN--S$_1$ (m1) was less effective than the other models (channels). 

\begin{table}[!b]
\begin{center}
\caption{Average STOI scores of the noisy utterances tests (input) and enhanced utterances (output) of the individual DNN--S channels.}\label{tab:stoinoidnns}
\begin{tabularx}{\columnwidth}{>{\centering}m{0.9cm}|>{\centering}m{0.61cm}>{\centering}m{0.61cm}>{\centering}m{0.61cm}>{\centering}m{0.61cm}>{\centering}m{0.61cm}>{\centering}m{0.61cm}>{\centering\arraybackslash}X}
\toprule
\hline
\textbf{STOI}&\textbf{m1}&\textbf{m2}&\textbf{m3}&\textbf{m4}&\textbf{m5}&\textbf{m6}&\textbf{m7}\\
\hline
\textbf{Noisy}&\textbf{0.672}&0.581&0.631&0.668&0.671&0.666&0.663\\
{\scriptsize\textbf{DNN--S}}&0.670&\textbf{0.677}&\textbf{0.668}&\textbf{0.685}&\textbf{0.695}&\textbf{0.679}&\textbf{0.673}\\
\hline
\bottomrule
\end{tabularx}\vspace{-0.1cm}
\end{center}
\end{table}
\begin{table}[!b]
\begin{center} 
\caption{Average SSNRI of enhanced utterances of individual channels of DNN--S.}\label{tab:sdinoidnns}
\begin{tabularx}{\columnwidth}{>{\centering}m{0.9cm}|>{\centering}m{0.61cm}>{\centering}m{0.61cm}>{\centering}m{0.61cm}>{\centering}m{0.61cm}>{\centering}m{0.61cm}>{\centering}m{0.61cm}>{\centering\arraybackslash}X}
\toprule
\hline
\textbf{SSNRI}&\textbf{m1}&\textbf{m2}&\textbf{m3}&\textbf{m4}&\textbf{m5}&\textbf{m6}&\textbf{m7}\\
\hline
\textbf{Noisy}&0.000&0.000&0.000&0.000&0.000&0.000&0.000\\
{\scriptsize\textbf{DNN--S}}&\textbf{8.722}&\textbf{13.179}&\textbf{12.859}&\textbf{12.132}&\textbf{8.699}&\textbf{12.530}&\textbf{12.127}\\
\hline
\bottomrule
\end{tabularx}\vspace{-0.1cm}
\end{center}
\end{table}
\begin{table}[!b]
\begin{center}
\caption{Average STOI and SSNRI scores of the enhanced utterances of DNN--S, DNN--F, and DNN--C under all noise conditions and microphone channels.}\label{tab:overall}
\begin{tabularx}{\columnwidth}{>{\centering}m{1cm}|>{\centering}m{1.9cm}>{\centering}m{1.9cm}>{\centering\arraybackslash}X}
\toprule
\hline
&\textbf{DNN--S}&\textbf{DNN--F}&\textbf{DNN--C}\\
\hline
\textbf{STOI}&0.678&0.764&\textbf{0.770}\\
\textbf{SSNRI}&11.464&12.760&\textbf{16.729}\\
\hline
\bottomrule
\end{tabularx}\vspace{-0.1cm}
\end{center}
\end{table}

In Table \ref{tab:overall}, the averaged STOI and SSNRI scores of the enhanced utterances of DNN--S, DNN--F, and DNN--C under all noise conditions and microphone channels are shown. For DNN--S, here we report the STOI and SSNRI scores averaged over all channels shown in Tables \ref{tab:stoinoidnns} and \ref{tab:sdinoidnns}. From these tables, it is clear that both DNN--F and DNN--C achieve superior evaluation scores than DNN--S, and these results further confirm that the two MC SE systems outperform DNN--S, a single-channel SE system, in improving the intelligibility and SNR of noisy utterances. Furthermore, both evaluation metrics indicate that DNN--C outperforms DNN--F, revealing the superiority of the two-stage SE architecture in promoting intelligibility and noise reduction for distorted signals.

\begin{figure}[!t]
{\centering \includegraphics[width=0.9\columnwidth]{\FIGUREPATH 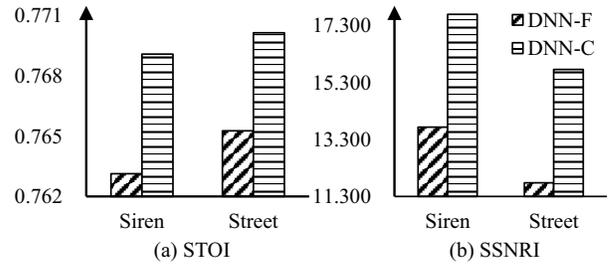}
\caption{The (a) STOI and (b) SSNRI scores of DNN--F and DNN--C enhanced noisy utterances in siren and street noise environmentss, with an average of more than three SNR levels.}\label{fig:detail}}
\end{figure}

Figure \ref{fig:detail} compares DNN--F and DNN--C under siren and street noise conditions under all SNR levels in terms of the averaged (a) STOI and (b) SSNRI evaluation metrics. From this figure, we further confirm that DNN--C provides higher intelligibility and more significant SNR improvements than DNN--F.

\begin{figure}[!t]
{\centering \includegraphics[width=0.9\columnwidth]{\FIGUREPATH 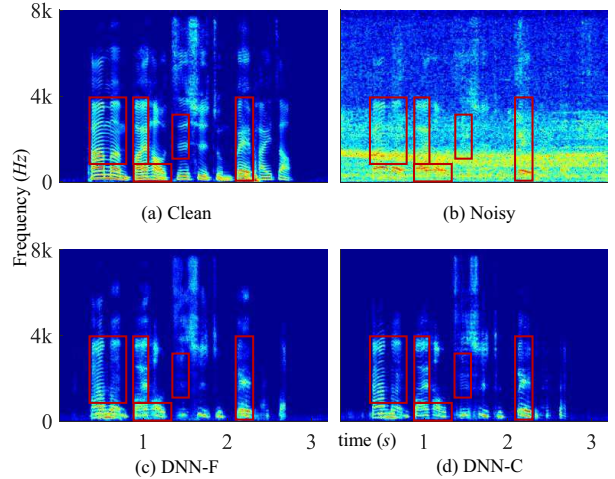}
\caption{Spectrograms of (a) a clean utterance, (b) the noisy signal recorded by m1, (c) the DNN--F enhanced speech, and (d) the DNN--C enhanced version.}\label{fig:specana}}
\end{figure}

Figures \ref{fig:specana}(a)-(d) show the spectrograms of an utterance under four conditions: (a) clean noise-free, (b) noise-corrupted (with a PESQ score of 1.642), (c) noise-corrupted and then enhanced by DNN--F (with a PESQ score of 1.834), and (d) noise-corrupted and then enhanced by DNN--C (with a PESQ score of 1.923). The utterance was corrupted with street noise at -5 dB SNR, and the noisy utterance was recorded by the m1 microphone. From these figures, we find that the spectrogram of the DNN--C-enhanced utterance in Fig. \ref{fig:specana}(d) is quite similar to that of the clean utterance in Fig. \ref{fig:specana}(a). In addition, by comparing Fig. \ref{fig:specana}(d) to Fig. \ref{fig:specana}(c), DNN--C reveals clearer spectral characteristics and sound structures than DNN--F, as indicated by the red blocks. These observations also explain why DNN--C achieved higher evaluation scores than DNN--F.

\section{Conclusion}\label{sec:conclu}
In this study, we presented three DNN--based SE systems under a distributed-microphone scenario applied in the diffuse-noise field environment. These three systems (DNN--S, DNN--F, and DNN--C) were evaluated in the TMHINT dataset. Experimental results showed that all of these systems were able to reduce the noise effect to improve speech intelligibility. Meanwhile, the two-stage DNN--C system achieved the optimal objective intelligibility score among the three systems. In the future, we plan to perform the DNN-based distributed-microphone SE task by properly selecting the recording channels rather than using them all as the system input. Furthermore, in addition to diffuse-noise environments, the DNN-C algorithm will be applied to reverberant environments to examine its capability in reducing the respective distortion.

\vfill\pagebreak

\bibliographystyle{ieeetr}
\bibliography{\RESOURCEPATH reference}

\end{document}